# Combining detection and reconstruction of correlational and quasi-periodic motifs in viral genomic sequences with transitional genome mapping: Application to COVID-19

Vladimir R. Chechetkin[1,]* and Vasily V. Lobzin

[1]Engelhardt Institute of Molecular Biology of Russian Academy of Sciences, Vavilov str., 32, Moscow 119334, Russia

*Corresponding author. E-mails chechet@eimb.ru; vladimir_chechet@mail.ru

[2]School of Physics, University of Sydney, Sydney, NSW 2006, Australia

E-mail vasili.lobzin@sydney.edu.au

**Abstract**

A method of Transitional Automorphic Mapping of the Genome on Itself (TAMGI) is aimed at combining detection and reconstruction of correlational and quasi-periodic motifs in the viral genomic RNA/DNA sequences. The motifs reconstructed by TAMGI are robust with respect to indels and point mutations and can be tried as putative therapeutic targets. We developed and tested the relevant theory and statistical criteria for TAMGI applications. The applications of TAMGI are illustrated by the study of motifs in the genomes of the severe acute respiratory syndrome coronaviruses SARS-CoV and SARS-CoV-2 (the latter coronavirus SARS-CoV-2 being responsible for the COVID-19 pandemic) packaged within filament-like helical capsid. Such ribonucleocapsid is transported into spherical membrane envelope with incorporated spike glycoproteins. Two other examples concern the genomes of viruses with icosahedral capsids, satellite tobacco mosaic virus (STMV) and bacteriophage PHIX174. A part of the quasi-periodic motifs in these viral genomes was evolved due to weakly specific cooperative interaction between genomic ssRNA/ssDNA and nucleocapsid proteins. The symmetry of the capsids leads to the natural selection of specific quasi-periodic motifs in the related genomic sequences. Generally, TAMGI provides a convenient tool for the study of numerous molecular mechanisms with participation of both quasi-periodic motifs and complete repeats, the genome organization, contextual analysis of *cis/trans* regulatory elements, data mining, and correlations in the genomic sequences.

# 1. Introduction

The development of antiviral drugs is based mainly on targeting specific motifs in viral proteins and/or viral genomes (De Clercq, 2006; Müller and Kräusslich, 2009; Lou et al., 2014; Kretova et al., 2017; Kravatsky et al., 2017). Sequencing of viral genomes provides the primary source of information (Wohl et al., 2016; Houldcroft et al., 2017). Then, the viral sequences are annotated against available databases (Sharma et al., 2016; Ibrahim et al., 2018). The choice of putative targets in viral genomes is strongly hampered by high frequency of point mutations and indels. In this paper we describe a method for combined detection and reconstruction of correlational and quasi-periodic motifs which is approximately robust with respect to point mutations and indels.

The method of Transitional Automorphic Mapping of the Genome on Itself (TAMGI) is based on the extension and adaptation of the well-known correlation function technique (see, e.g., Chechetkin and Turygin, 1996; Li, 1997; Lobzin and Chechetkin, 2000; Bernaola-Galván et al., 2002; and further references therein). The primary objects in our approach are correlational motifs. Generally, they cannot be reduced to the sparse or tandem repeats with gaps and alignments. As periodic features produce decaying or persistent oscillations in correlational motifs separated by distances multiple to the period, the periodic features can also be detected by the suggested method. TAMGI is able to reconstruct tandem repeats (both complete and incomplete) as well. In this latter problem it overlaps with the methods developed by the other authors (Benson, 1999; Szklarczyk and Heringa, 2004; Boeva et al., 2006; Sokol et al., 2007; Shi and Liang, 2019). With respect to the whole genome, TAMGI can be considered as an analog of principal component analysis. The distribution of correlational motifs over the genome provides information about large-scale genome organization.

The method of TAMGI was sketched briefly in our paper (Chechetkin and Lobzin, 2020) when applying to the study of ribonucleocapsid assembly/packaging signals in the genomes of the severe acute respiratory syndrome coronaviruses SARS-CoV and SARS-CoV-2. In this publication we develop the extended theory and derive the relevant statistical criteria for analysis of the output set of patterns in TAMGI. To illustrate the applications, we chose the genomes of the coronaviruses SARS-CoV and SARS-CoV-2 packaged within the helical capsid as well as the genomes of two viruses with the icosahedral capsids, satellite tobacco mosaic virus (STMV) and bacteriophage ϕX174 studied previously (Chechetkin and Lobzin, 2019, 2020) by different methods. Such choices allow us to cross-check the features detected by TAMGI and other methods.

# 2. Theory and methods

## *2.1. Transitional genome mapping*

The algorithm for TAMGI with a step *s* is defined as follows. Let a nucleotide $N_{m,\alpha}$ of the type α be positioned at a site *m* of the genomic sequence. Then, a pair of *s*-neighbors, $N_{m-s}$ and $N_{m+s}$, is searched for around $N_{m,α}$. The nucleotide $N_{m,α}$ will be retained if it has at least one *s*-neighbor $N_{m-s,α}$ or $N_{m+s,α}$ of the same type and be replaced by void otherwise (denoted traditionally by hyphen). All *s*-neighbors of the same type, $N_{m-s,α}$ or/and $N_{m+s,α}$, should also be retained. The resulting sequence after TAMGI is composed of the nucleotides of four types (A, C, G, T) and the hyphens "-" denoting voids. Further analysis is reduced to the study of all complete words of length *k* (*k*-mers) composed only of nucleotides (voids within the complete words are prohibited) and surrounded by the voids "-" at 5'- and 3'-ends, -$N_k$-. By definition, the complete words are non-overlapping. At the next stage, the mismatches with hyphens to the complete words can be studied.

To avoid end effects and to ensure homogeneity of the mapping, the linear genomes will always be circularized,

$$N_{m,α}^c = \begin{cases} N_{m,α}, & \text{if } 1 \leq m \leq M; \\ N_{m-M,α}, & \text{if } M+1 \leq m \leq 2M-1, \end{cases} \quad (1)$$

where $N_{m,α}$ denotes the nucleotide of the type α ∈ (A, C, G, T) positioned at the site *m* and *M* is the genome length. Choosing circularized mapping is natural and convenient for the study of quasi-repeating motifs. The theory and simulations show that for even *M* the step *M*/2 should be considered apart from the other steps. Therefore, the range of steps can be chosen from 1 to

$$L = \begin{cases} [M/2] \text{ for odd } M; \\ M/2 - 1 \text{ for even } M, \end{cases} \quad (2)$$

where the brackets denote the integer part of the quotient. Any sequence can be expanded via the complete set of TAMGI components (TAMGI sequences for particular steps) with the steps from *s*=1 to [*M*/2]. This means that TAMGI components can be considered as the generalized genome coordinates or the principal components related to the genome organization.

The circularized version of TAMGI can also be described as follows. (i) Take and circularize a linear genome. Superimpose two identical circular genomes over each other. (ii) Rotate clockwise one of the genomes on a step *s* and count all coincidences between two genomes. (iii) Rotate counterclockwise one of the genomes on a step *s* and count all coincidences between two genomes. (iv) Unite all coincidences into one sequence and fill the voids by hyphens.

We explain the algorithm using particular fragment of 20 nt at the start of the genome for the coronavirus SARS-CoV, 5'-ATATTAGGTTTTTACCTACC-3'. Let us choose the step *s*=3 for example. The nucleotide T at the site *m*=2 has the neighboring nucleotide T at the site *m*=5; both nucleotides should be retained. The nucleotide A at the site *m*=3 has the neighboring nucleotide A at the site *m*=6; both nucleotides should also be retained. The nucleotide T at the site *m*=4 has no T-neighbors at both sites *m*=1 and 7 and should be replaced by hyphen, etc. If this fragment is circularized, the nucleotide A at the site *m*=18 will have the neighboring nucleotide A at the site *m*=1 and both nucleotides should be retained.

Finally, the resulting sequence after TAMGI with the step $s=3$ has the form, ATA-TA--TT-TT--C-AC-, and contains 1-mer, -C-; 2-mers, -TA-, -TT-, -TT-, -AC-; and 3-mer, -ATA-. Let the fragment above be placed within region surrounded by indels, indel|ATATTAGGTTTTTACCTACC|indel. Then, the application of TAMGI with the step $s=3$ to such sequence retains all neighboring nucleotides within fragment, except possibly the nucleotides at the boundaries depending on the other neighbors. This means that TAMGI is robust with respect to indels if the step $s$ is less than the distances between indels. Let the length of the genome be $M$. If the step of TAMGI is $s$ and the number of indels is $N_{ID}$, they affect the correlation motifs in the region $2sN_{ID}$. The fraction of modified motifs is $2sN_{ID}/M$. This imposes the restriction $s < M/2N_{ID}$ for the approximate conservation of primary motifs in the presence of indels. In the range $s > M/2N_{ID}$ TAMGI can be applied to the study of the general character of correlations which remains robust in the presence of indels. Therefore, the duality of the combined detection/reconstruction analysis by TAMGI provides additional opportunities and covers all range of the steps from 1 to $L$ defined by Eq. (2). The application of TAMGI to a sequence with complete tandem repeats yields long words composed of repeats if the step $s$ coincides with the length of repeats.

The typical protocol for application of TAMGI to the study of the viral genomic sequences is as follows. Take conventional reference sequence from GenBank and perform the complete TAMGI analysis for such sequence. Then, using isolates with real point mutations and indels, assess the conservation of motifs and variations in correlations obtained by TAMGI. The most conserved motifs can be recommended as putative therapeutic targets related to medical applications. Using real sequencing data for the assessment of mutation impact on motifs is essential because the rate of mutations strongly varies for different viruses. Many mutations make the virus unviable and lead to its extinction from population. Only neutral or compatible mutations are permissible. Some rare mutations can be considered as favorable. The mutations may be distributed over viral genomes strongly inhomogeneously and there are conserved (approximately or strictly) regions on viral genomes with small frequency of mutations (see, e.g., Kretova et al., 2017; Kravatsky et al., 2017). These complicated effects cannot be assessed by simplified simulations.

*2.2. Correlation motifs and tandem repeats*

Unlike repeats (tandem, sparse, complete and incomplete), which are quite common objects in genetics, the correlation motifs seem to be not considered before. In this paper, the correlation motifs are defined as a set of $k$-mers generated by TAMGI. Therefore, after presentation of TAMGI algorithm, it would be useful to compare tandem and correlation motifs. Consider, e.g., the sequence ATGATCGGC. If the repeats are searched for by a typical algorithm for triplets with one gap, one obtains AT-AT----, whereas TAMGI with $s = 3$ yields AT-ATC--C. In the case of tandem repeats, TAMGI can be reduced to the algorithm for incomplete tandem repeats after proper redefinitions and filtering, but generally the results are different.

The tandem repeats in the viral genomes are rarely encountered but occur sometimes. In particular, the genome of human coronavirus HCoV-HKU1 (GenBank accession: NC_006577.2) contains fragment with 14 tandem repeats of 30 nt, AATGACGATGAAGATGTTGTTACTGGTGAC, coding for amino acids NDDEDVVTGD (Woo et al., 2006). The application of TAMGI with the step $s = 30$ to this fragment provides long 420-mer composed of tandem repeats, whereas, e.g., the mapping with $s = 54$ yields correlation motifs ATGACGATGA repeating with spacing of 30 nt. As the period $p = 54$ is equal to the ribonucleocapsid helix pitch (see Section 3.1), this means that such correlation motifs may facilitate the encapsidation. Generally, long tandem repeats are simultaneously a source of a variety of correlation motifs which can play different functional roles and participate in various molecular mechanisms. Such correlation motifs modified by mutations can subsequently be scattered over the genome. Being more general object, the correlation motifs include tandem repeats as a particular case.

*2.3. Normalization of nucleotide frequencies after TAMGI*

The frequencies of nucleotides after TAMGI with the step $s$,

$$\varphi_{\alpha, s} = N_{\alpha, s} / M; \quad \varphi_{total, s} = \sum_{\alpha} \varphi_{\alpha, s}, \tag{3}$$

should be properly normalized to assess their statistical significance. The normalization ought to be performed against the counterpart characteristics in the random sequences of the same nucleotide composition. Below, we will always imply that the length of the genome satisfies the condition $M \gg 1$. The frequencies for the combinations

(non-$N_\alpha$)($N$)$_{s-1}N_\alpha(N)_{s-1}N_\alpha(N)_{s-1}$... $N_\alpha(N)_{s-1}N_\alpha (N)_{s-1}$(non-$N_\alpha$) (where $N_\alpha$ denotes the nucleotides of the type α, $N$ means any nucleotide and it is implied that there are $n$ nucleotides of the type α with the proper spacing $s$ within this combination) can be assessed in the random sequences as

$$p_{n, \alpha} = (1 - \varphi_\alpha)^2 \varphi_\alpha^n; \quad n = 1, 2, ..., \tag{4}$$

where $\varphi_\alpha$ is the frequency of nucleotides of the type α in the genome. Then, the frequency of the nucleotides of the type α after application of TAMGI, $\Phi_\alpha$, is expressed through the frequencies defined by Eq. (4) as follows,

$$\sum_{n=2}^{\infty} n p_{n, \alpha} = \Phi_\alpha = \varphi_\alpha^2 (2 - \varphi_\alpha), \tag{5}$$

while the total frequency of nucleotides after TAMGI is obtained by summation over $\Phi_\alpha$,

$$\Phi_{total} = \sum_{\alpha} \Phi_\alpha. \tag{6}$$

The variances for the frequencies defined by Eqs. (5) and (6) are determined for the random sequences by the binomial distribution,

$$\sigma^2(\Phi_\alpha) = \Phi_\alpha(1-\Phi_\alpha)/M; \quad \sigma^2_{total} = \sum_\alpha \sigma^2(\Phi_\alpha). \tag{7}$$

This means that the total frequency defined by Eq. (6) can be presented in terms of a normalized deviation,

$$\kappa_s = (\varphi_{total,s} - \Phi_{total})/\sigma_{total}, \tag{8}$$

(for notations, see Eqs. (3) and (5)–(7)). As can be proved (see also below), the deviations (8) in the random sequences are governed by the Gaussian statistics and the deviations for the different steps $s$ are approximately independent. The steps with the pronounced deviations of high statistical significance (see Eq. (8) and Section 2.5 below) are of primary interest. For quasi-periodic motifs with a period $p$, the deviations (8) should reveal approximately equidistant peaks at the steps $s = p, 2p, ...$ A part of statistically significant deviations may be associated with correlations between nucleotides. The normalized deviations defined by Eq. (8) unify dynamical range and presentation of data. Such presentation facilitates the comparison of genomes with different nucleotide composition.

*2.4. Distribution of k-mers after TAMGI*

The frequency of words of length $k$, $-N_k-$ (where $N$ means nucleotide of any type), in the random sequences can be assessed as,

$$p_k = (1-\Phi_{total})^2 \Phi_{total}^k; \quad k = 1, 2, ..., \tag{9}$$

where $\Phi_{total}$ is defined by Eqs. (5) and (6). Let there be $n_{k,s}$ words of length $k$ after TAMGI with the step $s$. Then, the empirical probability to detect a word longer than $k'$ can be determined as,

$$\tilde{\Pr}_s(k \geq k') = \sum_{k=k'}^{k_{max}} n_{k,s} / \sum_{k=1}^{k_{max}} n_{k,s}. \tag{10}$$

The summation in Eq. (10) is performed up to the maximum detected length. The frequency of words with lengths exceeding $k'$ in the random sequences is given by

$$\Phi(k \geq k') = \sum_{k=k'}^{\infty} p_k = (1-\Phi_{total})\Phi_{total}^{k'}. \tag{11}$$

The probability to find a word with a length exceeding $k'$ in the complete pool of words can be calculated as

$$\Pr(k \geq k') = \Phi(k \geq k')/\Phi(k \geq 1) = \Phi_{total}^{k'-1}. \tag{12}$$

The empirical probability for the natural genomic sequences (10) should be compared with the predictions for the random sequences expressed by Eq. (12). The difference between natural and theoretical distributions can be statistically assessed by the Kolmogorov-Smirnov criterion.

The probability density corresponding to probability (12) is defined as

$$p(k) = \Phi_{total}^{k-1} - \Phi_{total}^{k}; \quad k = 1, 2, \ldots \quad (13)$$

Then, the mean length of words and its variance in the random sequences after TAMGI are given by

$$<k> = \sum_{k=1}^{\infty} k p(k) = 1/(1-\Phi_{total}), \quad (14)$$

$$\sigma^2(k) = <k^2> - <k>^2 = \Phi_{total}/(1-\Phi_{total})^2. \quad (15)$$

The mean number of all words in the random sequences after TAMGI can be assessed as

$$n_{words} = M\,\Phi(k \geq 1) = M\,\Phi_{total}(1-\Phi_{total}) = M\,\Phi_{total}/<k>. \quad (16)$$

The maximum length of words within the interval of steps $\Delta s = s_2 - s_1 + 1$ in the random sequences can be estimated using Bonferroni approximation as

$$M(s_2 - s_1 + 1)(1-\Phi_{total})\Phi_{total}^{k'_{max,\Delta s}} = N_{thr}. \quad (17)$$

A typical threshold for the statistical significance is $N_{thr} = 0.05$, whereas the lower boundary for the maximum length corresponds to $N_{thr} \cong 1$.

## 2.5. Tests

The theoretical predictions for TAMGI of random sequences were tested by reshuffling the genomic sequence for SARS-CoV ($M = 29751$, $N_A = 8481$, $N_G = 6187$, $N_T = 9143$, $N_C = 5940$). The frequencies averaged over 10,000 random realizations were in complete agreement with the theoretical results (5)–(7). The spectrum for normalized deviations (8) is shown in Fig. 1A and is homogeneous for the circularized genome. As expected, the normalized deviations (8) are governed by Gaussian statistics (see Fig. 1B). The distribution of $k$-mer lengths after TAMGI for a particular arbitrarily chosen step is shown in Fig. 1C, while Fig. 1D shows the distribution of $k$-mer lengths for the steps 1–500. Both in Figs. 1C and 1D the deviations for distributions of word lengths for particular random realization from theoretical probability defined by Eq. (12) appeared to be insignificant by Kolmogorov-Smirnov criterion. As expected from Eq. (17), the maximum length detected for the set associated with the interval 1–500 appeared to be longer in comparison with that for a particular step. Thus, the tests prove that suggested characteristics can be used for the study of real genomes.

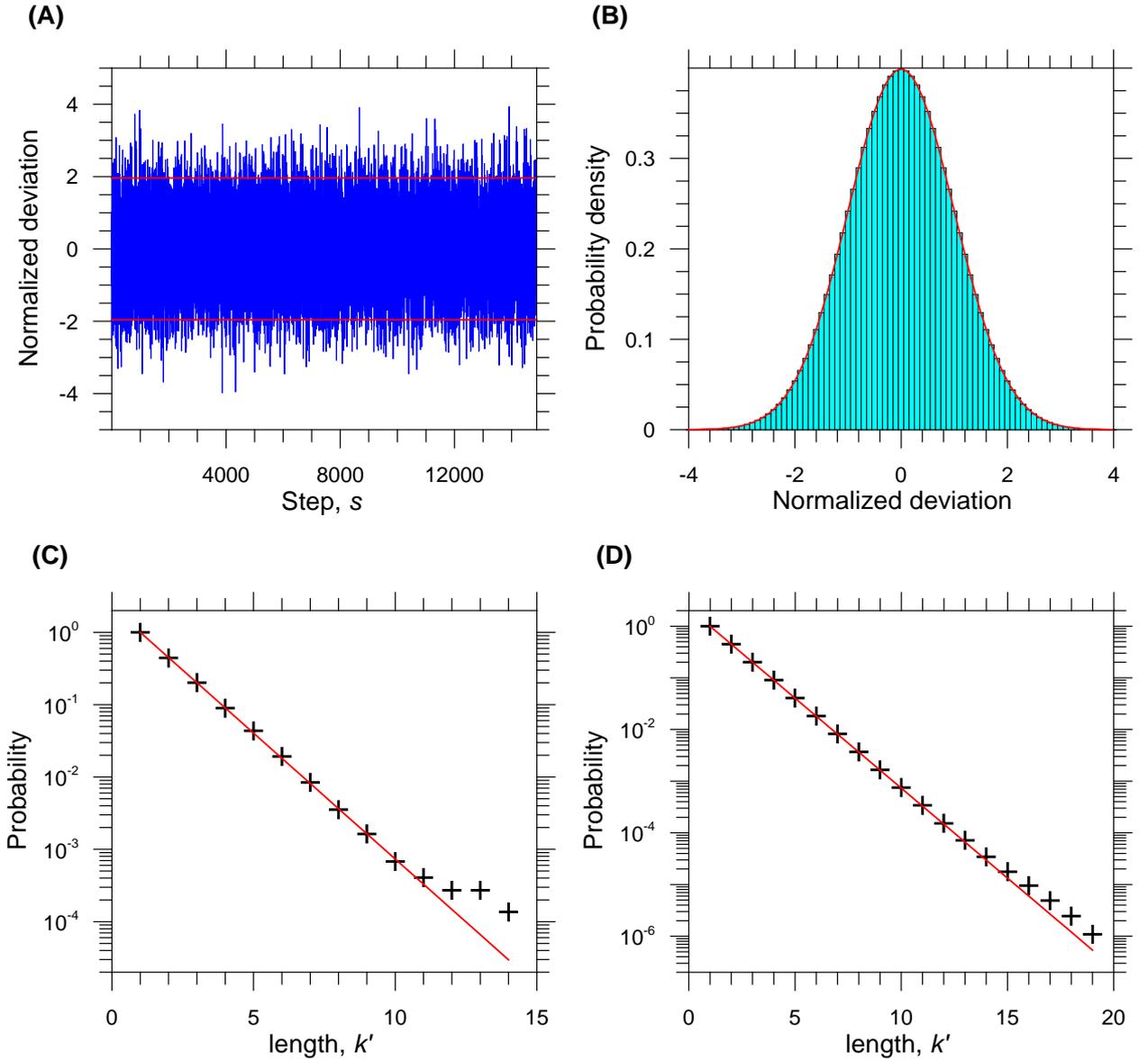

**Fig. 1.** (**A**) The normalized deviations defined by Eq. (11) for a particular random realization obtained by reshuffling genomic sequence for SARS-CoV ($M = 29751$, $N_A = 8481$, $N_G = 6187$, $N_T = 9143$, $N_C = 5940$). The horizontal lines correspond to the Gaussian $p$-value 0.05. (**B**) The distribution of the normalized deviations shown in panel A and its comparison with the standard Gaussian distribution $N(0, 1)$. (**C**) The distribution of $k$-mer lengths after TAMGI for a random sequence obtained by reshuffling genomic sequence for SARS-CoV for a particular arbitrarily chosen step and its comparison with the theoretical prediction based on Eq. (12) (straight line). (**D**) The distribution of $k$-mer lengths after TAMGI for a random sequence obtained by reshuffling genomic sequence for SARS-CoV for the steps within interval 1–500 and its comparison with the theoretical prediction based on Eq. (12) (straight line).

## 3. Results

For illustration of TAMGI applications, we chose motifs in viral genomes evolved due to specific interactions between genomic RNA/DNA during packaging into protein capsid (Rossmann and Rao, 2012; Mateu, 2013). As the main information in viral genomes is related to the coding for different proteins needed for virus proliferation, the signals corresponding to genome packaging are evolved using the redundancy of the genetic code. The detection and reconstruction of such motifs are especially challenging because the specificity of molecular interactions is rather weak, while the point mutations and indels induced mainly during replication stage are frequent. Helical, icosahedral or prolate capsids of viruses possess a certain symmetry which can serve as a guiding thread to search for segment regularities and distribution of motifs in the viral genomes (Chechetkin and Lobzin, 2019, 2021).

*3.1. Quasi-periodic motifs in the genomes of the coronaviruses SARS-CoV and SARS-CoV-2*

At the beginning of April 2021, the number of coronavirus disease 19 (COVID-19) cases exceeded 131.2 million, while the number of deaths over all world exceeded 2.9 million (https://www.worldometers.info/coronavirus/). The pandemic is still continuing and the second and third waves of diseases are pending in many countries. The development of efficient medications and vaccines against the coronaviruses needs the knowledge of main molecular mechanisms in the virus life cycle and virus-host interaction (Ziebuhr, 2016; Saxena, 2020; O'Leary et al., 2020; Mishra and Tripathi, 2021). The long (about 30,000 nt) non-segmented plus-sense single-stranded RNA genome of the coronaviruses is packaged within a filament-like helical nucleocapsid, while the whole ribonucleocapsid is packaged within a membrane envelope with spike glycoproteins in a mode resembling the outlines of the flower petals (Neuman and Buchmeier, 2016; Masters, 2019; Chang et al., 2014; Chen et al., 2007; Gui et al., 2017). The coronavirus SARS-CoV-2 caused the outbreak of COVID-19 pandemic and appeared to be much more virulent than its relative SARS-CoV. The cryogenic electron microscopy (cryo-EM) have revealed that the ribonucleocapsid of SARS-CoV is helical with an outer diameter of 16 nm and an inner diameter of 4 nm (Chang et al., 2014). The turn of the nucleocapsid is composed of two octamers polymerized from dimeric N proteins (Chen et al., 2007). The pitch for the SARS-CoV nucleocapsid is 14 nm. The packaging of the SARS-CoV ssRNA genome near internal surface of helical nucleocapsid with such parameters should correspond to 54–56 nt per helical turn (or 6.75–7 nt per N protein) (Chang et al., 2014; Chechetkin and Lobzin, 2020). The counterpart cryo-EM data for SARS-CoV-2 are yet absent. The bioinformatic analysis may elucidate similarities between structural characteristics of the nucleocapsids for two coronaviruses. The whole ribonucleocapsid structure of coronaviruses remains invariant under transition by one helical turn. Due to the transitional symmetry of a helix, weakly specific cooperative interaction between ssRNA and nucleocapsid proteins should lead to the natural selection of specific quasi-periodic motifs in the related genomic sequences (Chechetkin and Lobzin, 2020). In this section we will show how these features can be established with TAMGI method.

We took for comparative analysis and illustration the reference genomic sequence for SARS-CoV (GenBank accession: NC_004718; $M = 29751$, $N_A = 8481$, $N_G = 6187$, $N_T = 9143$, $N_C = 5940$) and the genomic sequence for one of isolates of SARS-CoV-2 (GenBank accession: MT371038; $M = 29719$, $N_A = 8873$, $N_G = 5834$, $N_T = 9554$, $N_C = 5458$). The general overviews for the corresponding normalized deviations defined by Eq. (8) for TAMGI steps from 1 up to $[M/2]$ are shown in Figs. 2A and 2B. About two-thirds of the genome for both viruses code for replicase proteins. The large-scale segmentation related to such coding mode is clearly seen in Figs. 2A and 2B. The pronounced peaks corresponding to quasi-periodic motifs with period $p \approx 54$ are shown in the inserts. The deviation for the step $s = 54$ appeared to be the highest for TAMGI data for SARS-CoV, whereas the deviation for the step $s = 54$ is a bit lower than the highest deviation for $s = 9$ for TAMGI data for SARS-CoV-2. The deviations for the multiples of the step $s = 54$ are also significant and indicate its quasi-periodic nature.

The distributions of $k$-mer lengths after TAMGI with the step $s = 54$ are shown in Figs. 2C and 2D and compared with the counterpart distributions for particular random realizations. The corresponding distributions of $k$-mer lengths after TAMGI with the steps within the interval $s = 1–500$ are shown in Figs. 2E and 2F and also compared with the counterpart distributions for particular random realizations. The deviations between all natural and random distributions were highly statistically significant by Kolmogorov-Smirnov criterion (Pr $< 10^{-5}$).

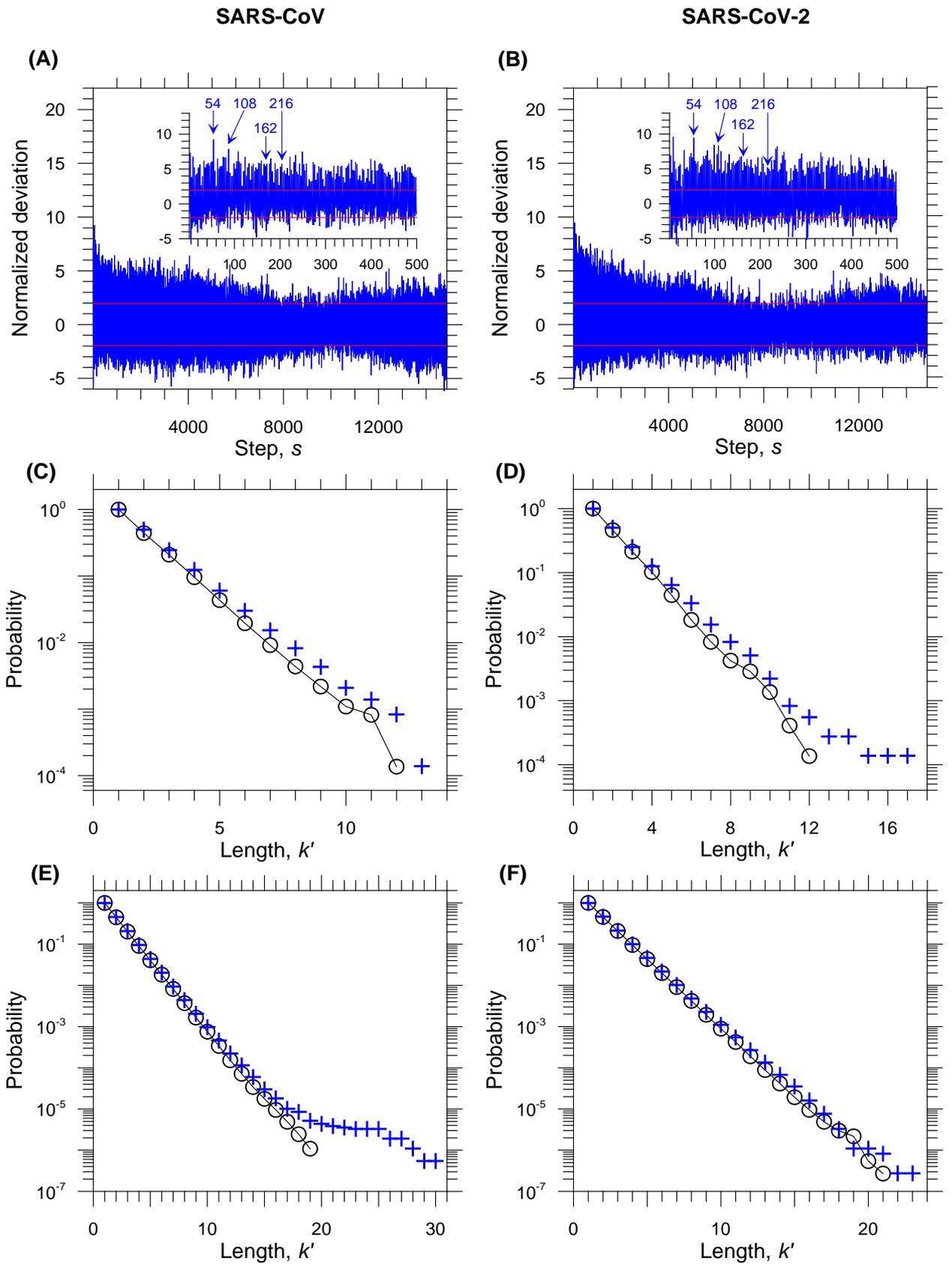

**Fig. 2. (A)** The normalized deviations defined by Eq. (8) after application of TAMGI to the genomic sequence for SARS-CoV (GenBank accession: NC_004718). The insert shows the peak deviation at the step $s = 54$ associated with the packaging signals related to the packaging of ssRNA genome into the helical capsid. The peaks at the multiple steps for $s = 54$ indicate the quasi-periodic nature of these signals. The horizontal lines correspond to the Gaussian $p$-value 0.05 for the random sequences. **(B)** The normalized deviations defined by Eq. (8) after application

of TAMGI to the genomic sequence for SARS-CoV-2 (GenBank accession: MT371038). The insert shows the peak deviation at the step $s = 54$ associated with the packaging signals related to the packaging of ssRNA genome into the helical capsid. The peaks at the multiple steps for $s = 54$ indicate the quasi-periodic nature of these signals. The horizontal lines correspond to the Gaussian *p*-value 0.05 for the random sequences. **(C)** The distribution of *k*-mer lengths after TAMGI with the step $s = 54$ for the genome of SARS-CoV (shown by crosses) and its comparison with the counterpart distribution for a random reshuffled sequence (shown by circles). **(D)** The distribution of *k*-mer lengths after TAMGI with the step $s = 54$ for the genome of SARS-CoV-2 (shown by crosses) and its comparison with the counterpart distribution for a random reshuffled sequence (shown by circles). **(E)** The distribution of *k*-mer lengths after TAMGI for the steps within interval 1–500 for the genome of SARS-CoV (shown by crosses) and its comparison with the counterpart distribution for a random reshuffled sequence (shown by circles). **(F)** The distribution of *k*-mer lengths after TAMGI for the steps within interval 1–500 for the genome of SARS-CoV-2 (shown by crosses) and its comparison with the counterpart distribution for a random reshuffled sequence (shown by circles).

The sequences after TAMGI with the step $s = 54$ and the lists of the words, $-N_k-$, with $k \geq 12$ for TAMGI with the steps within interval $s = 1$–500 are explicitly reproduced in Supplements S1–S4. As is seen from the list of motifs in Supplement S3, a part of the longest words for SARS-CoV was generated by poly-A signal positioned at 3'-end of the genome and corresponding to transcription termination. The counterpart signal was not reproduced in the version of the genomic sequence for SARS-CoV-2. We retained, however, poly-A signal to illustrate the ability of TAMGI to detect such signals as well. By the estimates in Section 2.1, the motifs corresponding to the steps $s \leq 500$ should be robust in the presence of about 60 indels, whereas their number is commonly within 1–5. The comparison of repertoires of words with the lengths $k \geq 6$ reveals their partial divergence between SARS-CoV and SARS-CoV-2; however, such words appeared to be closely conserved for different isolates of SARS-CoV-2 despite the load from point mutations and indels (Chechetkin and Lobzin, 2020). The choice of the most conserved motifs at the step $s = 54$ and the epitopes in N proteins responsible for the interaction between ssRNA and N proteins provides the promising therapeutic targets for the development of an antiviral vaccine (Kramps and Elbers, 2017; Chang et al., 2016; Dutta et al., 2020) (see also (Chechetkin and Lobzin, 2020) for discussion and further references).

*3.2. Motifs in the genome of the satellite tobacco mosaic virus (STMV)*

The two next examples concern the viruses with icosahedral capsids. In the virus world, more than a half of viruses belong to such species (Rossmann and Rao, 2012; Mateu, 2013). The icosahedral symmetry comprises 15 axes of the second order, 10 axes of the third order, and 6 axes of the fifth order. The total number of operations for the icosahedral symmetry is 60. The correspondence should be searched between the (generally multiple) elements of icosahedral symmetry and the character of large-scale quasi-periodic segmentation induced by weakly specific cooperative interactions between genomic RNA/DNA and capsid proteins.

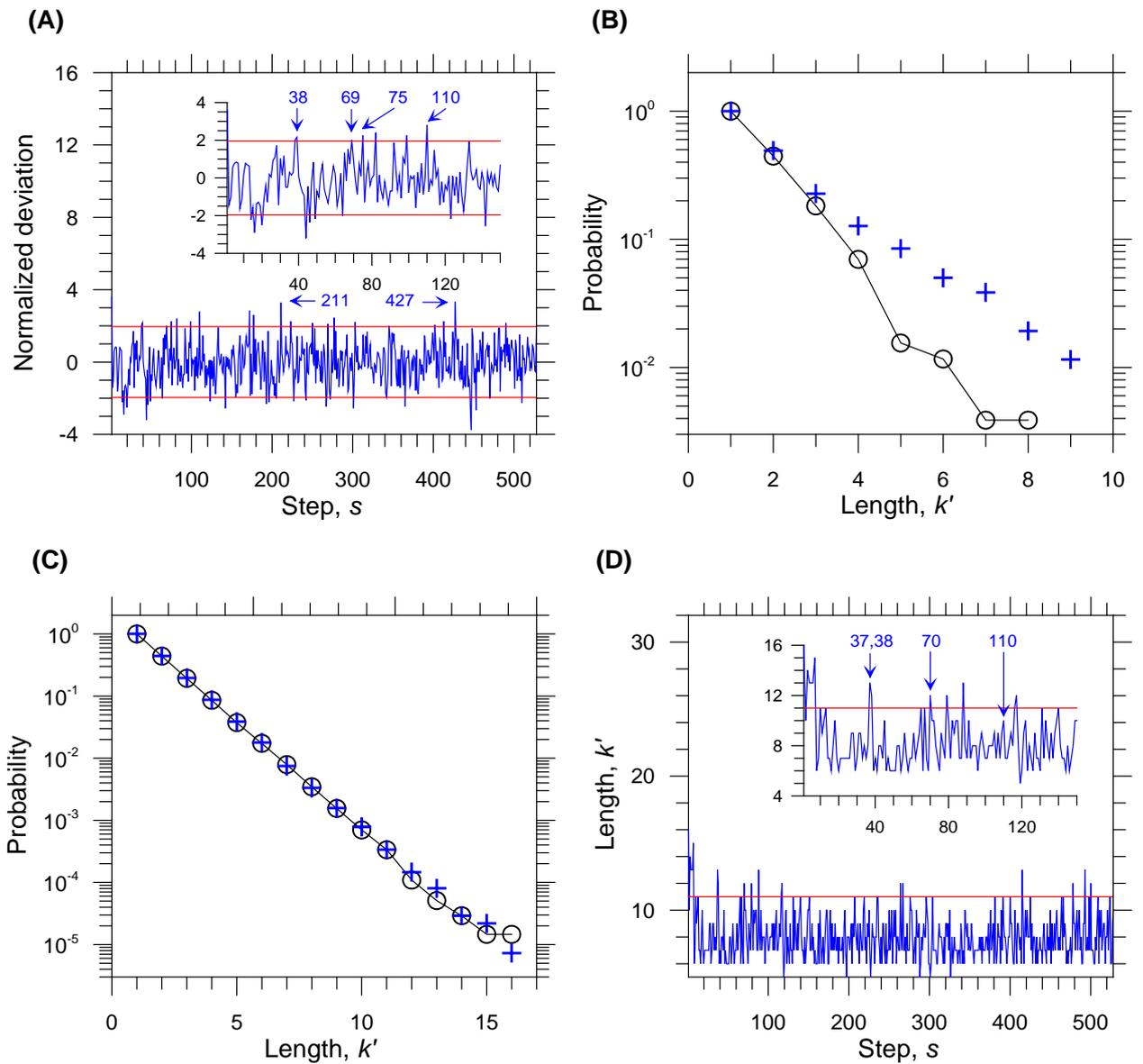

**Fig. 3. (A)** The normalized deviations defined by Eq. (8) after application of TAMGI to the genomic sequence for STMV (GenBank accession: M25782). The characteristic deviations at the steps associated with the segmentation related to the packaging of the genome into the icosahedral capsid are shown by arrows. The horizontal lines correspond to the Gaussian $p$-value 0.05 for the random sequences. **(B)** The distribution of $k$-mer lengths after TAMGI with the step $s = 211$ for the genome of STMV (shown by crosses) and its comparison with the counterpart distribution for a random reshuffled sequence (shown by circles). **(C)** The distribution of $k$-mer lengths after TAMGI for the steps from 1 to 528 for the genome of STMV (shown by crosses) and its comparison with the counterpart distribution for a random reshuffled sequence (shown by circles). **(D)** The plot for the maximum length of words at the step $s$. The horizontal lines correspond to the choice of parameters $\Delta s = 1$ and $N_{thr} = 0.05$ in Eq. (17).

STMV, a small icosahedral plant virus with linear positive-strand ssRNA genome, may be considered as one of the smallest reproducing species in nature (for a review and further references see (Dodds, 1998; Larson and McPherson, 2001)). Its reproducibility needs both a host cell and a host virus (tobacco mosaic virus in this case). The icosahedral capsid consisting of 60 identical subunits with genomic ssRNA inside was resolved on 1.4 Å scale (Larson et al., 2014). The visible RNA revealed 30 double-helical segments, each about 9 bp in length, packaged along the edges of capsid icosahedron (Zeng et al., 2012; Larson et al., 2014). The corresponding quasi-periodic segmentation on 30 segments with the period $p \approx 35.3$ nt was clearly pronounced in the Fourier spectra for this genomic sequence (Chechetkin and Lobzin, 2019).

The genomic sequence of STMV is of length $M = 1058$ (GenBank accession: M25782). About a half of the genome contains two overlapping ORF, the longer of which codes for coat protein, whereas the other half contains UTR. The overview of TAMGI deviations for the genome of STMV is shown in Fig. 3A. A significant deviation for $s = 38$ which can be associated with the period $p \approx 35.3$ nt appeared to be biased. The approximately equidistant peaks at $s = 75$ and $110$ were biased as well. Only a significant deviation at the step $s = 69$ can be associated with the doubled period $p \approx 35.3$ nt. The capsid assembly is presumed to be performed hierarchically via 5-fold intermediate (Rossmann et al., 1983). As genomic ssRNA participates actively in virion assembly (Patel et al., 2017), a similar mechanism may be suggested for RNA packaging. The peaks that can be associated with 5-fold segmentation are shown in Fig. 3A ($s = 211$ and $427$). The distribution of $k$-mers after TAMGI with the step $s = 211$ is presented in Fig. 3B, whereas the complete distribution of $k$-mers after TAMGI with the steps from 1 to 528 is shown in Fig. 3C. The deviations between these distributions and the counterpart distributions for random sequences were found to be statistically insignificant by Kolmogorov-Smirnov criterion. The sequence after TAMGI with $s = 211$ in GenBank format and the list of words with lengths $k \geq 9$ are reproduced explicitly in Supplements S5 and S6, respectively. The plot for the maximum length of words at the step $s$ is shown in Fig. 3D. The insert to Fig. 3D reveals short-range (up to $s = 10$–$20$) correlations for such steps; the shorter the step $s$, the longer the maximum length of word. Note also the correlations between significant deviation for the step $s = 38$ shown in Fig. 3A and the maximum $k$-mer lengths for $s = 37$ and $38$ ($k' = 13$ and $12$, respectively) as well as the other significant deviations and maximum lengths shown in Fig. 3D.

*3.3. Motifs in the genome of the bacteriophage ϕX174*

The bacteriophage ϕX174 belonging to *Microviridae* family presents an example of icosahedral viruses with ssDNA genome packaging (for a review see, e.g., Doore and Fane, 2016). The capsid of the mature virus is composed of 60 copies each of the coat protein F, the spike protein G, the DNA-binding protein J, and 12 copies of the pilot protein H (McKenna et al., 1994). The proteins H and J are associated with the inner capsid surface; the proteins H are beneath the centers of pentamers formed by the spike proteins G. There is one-to-one association between J and F proteins. The interactions with J proteins tether ssDNA to the inner capsid surface.

The circular genome of ϕX174 is of length $M = 5386$ and encodes 11 genes (GenBank accession: NC_001422). Fourier analysis of this genome revealed a large-scale segmentation compatible with the icosahedral symmetry (Chechetkin and Lobzin, 2019). The general overview for TAMGI deviations determined for the ϕX174 genome is shown in Fig. 4A. The highest deviation for TAMGI corresponds to $s = 180$ marked in the insert and can be associated with the segmentation on 30 segments compatible with icosahedral symmetry. The deviation for the approximately doubled step $s = 357$ was also significant, indicating a quasi-periodic nature of such segmentation. Another series of peaked deviations corresponds to the step $s = 30$ associated with 180-segmentation. The commensurability of the steps $s = 30$ and 180 indicates the hierarchical nature of the genome packaging. Note that both modes of segmentation on 30 and 180 segments were highly significant by Fourier analysis (Chechetkin and Lobzin, 2019). The deviation at the step $s = 90$ associated with segmentation on 60 segments, though not prominent, was also significant. The high deviation for $s = 1828$ can be associated with 3-fold segmentation.

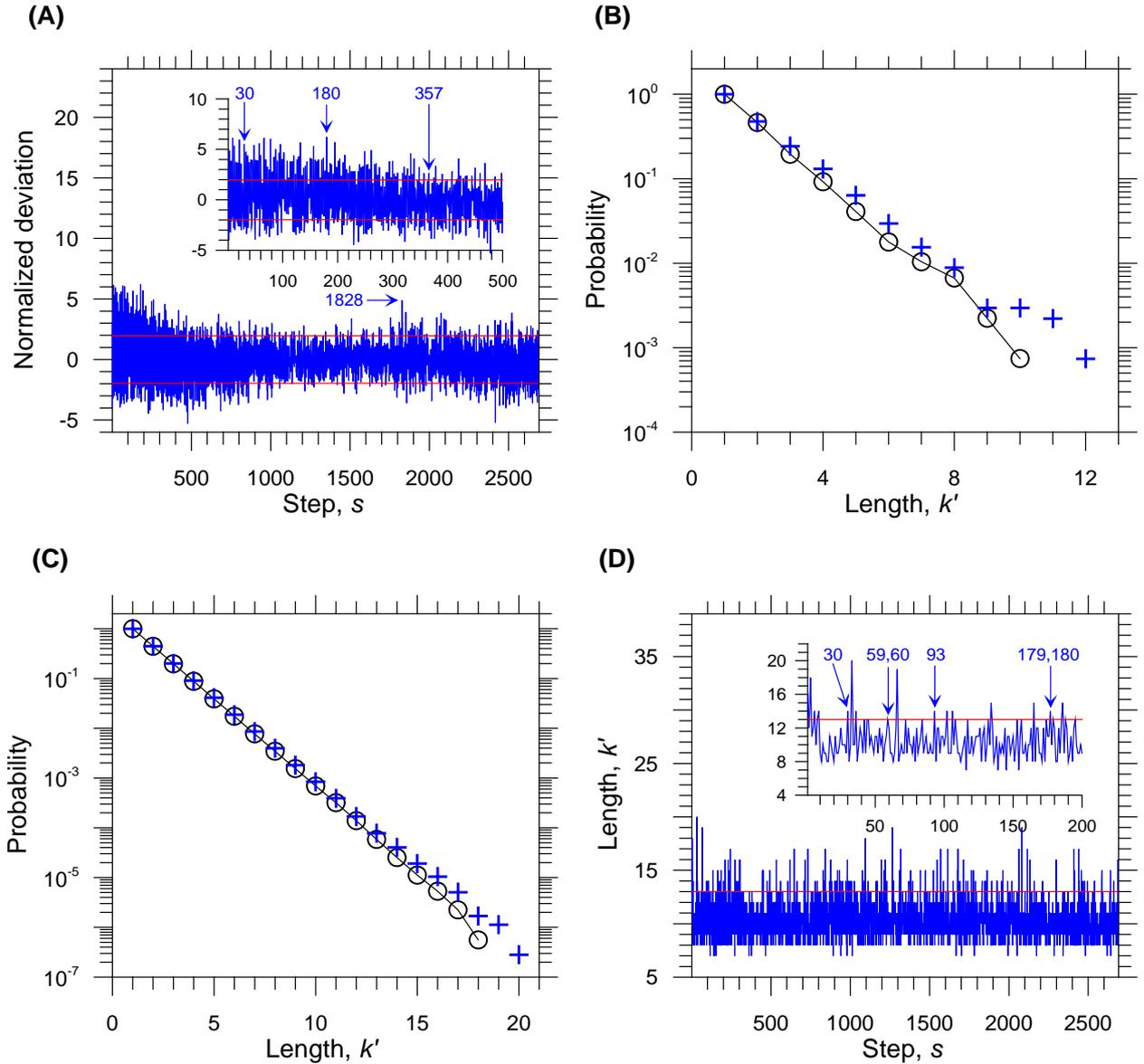

**Fig. 4. (A)** The normalized deviations defined by Eq. (8) after application of TAMGI to the genomic sequence for φX174 (GenBank accession: NC_001422). The characteristic deviations at the steps associated with the segmentation related to the packaging of the genome into the icosahedral capsid are shown by arrows. The horizontal lines correspond to the Gaussian $p$-value 0.05 for the random sequences. **(B)** The distribution of $k$-mer lengths after TAMGI with the step $s = 180$ for the genome of φX174 (shown by crosses) and its comparison with the counterpart distribution for a random reshuffled sequence (shown by circles). **(C)** The distribution of $k$-mer lengths after TAMGI for the steps from 1 to 2692 for the genome of φX174 (shown by crosses) and its comparison with the counterpart distribution for a random reshuffled sequence (shown by circles). **(D)** The plot for the maximum length of words at the step $s$. The horizontal lines correspond to the choice of parameters $\Delta s = 1$ and $N_{thr} = 0.05$ in Eq. (17).

The distributions of *k*-mers for the step *s* = 180 and for the steps from 1 to 2692 are presented in Figs. 4B and 4C in comparison with the counterpart distributions for the deviations for particular random realization. Though the differences between two distributions are poorly seen on the logarithmic scale, they are significant by the Kolmogorov-Smirnov criterion (Pr < $10^{-4}$). The plot for the maximum *k*-mer lengths at the step *s* is shown in Fig. 4D and again revealed short-range correlations seen in the insert. The clustering of maximum lengths around the steps associated with the characteristic segmentation modes compatible with icosahedral packaging is marked explicitly in Fig. 4D. The sequence after TAMGI with the step *s* = 180 in GenBank format and the list of words with lengths $k \geq 11$ for the steps from 1 to 2692 are presented in Supplements S7 and S8. Generally, the results of analysis by TAMGI and Fourier methods appeared to be concordant for the ϕX174 genome. The motifs found by TAMGI can be useful for the experimental study of the genome packaging.

## 4. Discussion and conclusions

The common paradigm for the cooperative specific interaction between genomic RNA/DNA and proteins is based on the consensus motifs in genomic sequences recognized by protein epitopes. The length of consensus motifs is implied to be approximately fixed. Generally, the motifs may be clustered into several groups with diverging consensus motifs. TAMGI indicates the possibility of the other scenario for the molecular evolution of motifs. In the molecular mechanisms related to the correlational or quasi-periodic motifs, the evolvement of motifs may start from approximately phased nucleotides of particular type followed by subsequent generation of the longer and longer motifs with a distribution resembling that of defined by Eq. (12). TAMGI reconstructs the longer motifs from the shorter ones. The abundance of primary phased nucleotides facilitates the generation of the longer motifs in comparison with counterpart motifs in the random sequences. As was argued above (Section 2.1), the correlational motifs are robust with respect to indels. Longer correlational motifs are rather rare (cf. Eqs. (9) and (12)) and occupy relatively small fraction of the genome (the longer the motifs, the smaller the fraction of the genome). If mutations in viral genomes were neutral, the occurrence of mutations directly within the motifs would be rare merely because of their small statistical weight. This means that correlational motifs are approximately robust to both point mutations and indels. The specific (correlational) surrounding of motifs and their robustness with respect to mutations makes them the plausible candidates on the functional significance in the mechanisms of virus life cycle. Indeed, the motifs that can be associated with packaging of viral genomes within capsid are persistently abundant in the genomes considered in Section 3. The longest motifs may serve as a seed of a particular molecular mechanism or play multi-functional role similar to *cis/trans* regulatory elements. The contextual analysis of the known *cis/trans* elements with TAMGI may elucidate the mechanisms of their evolving and be useful for data mining. The screening of the longest words against available databases would also be useful. The incorporation of relatively long (about 20 nt) correlation motifs into oligonucleotides immobilized on the surface of microarrays may facilitate the detection of viruses by microarrays (for a review on microarrays see, e.g., Dufva, 2009).

The extension of TAMGI on the mapping of complementary stretches is non-trivial. Such analysis is important, e.g., for the predictions of secondary structures of ssRNA viruses within the icosahedral capsids (Schroeder, 2014, 2018; Larman et al., 2017). The geometrical constraints imposed by capsid and the interaction between RNA and capsid proteins make such secondary structure different from free structure in solution. A possible extension of TAMGI for this problem can be achieved by using two windows of equal lengths $w$ with centers separated by step $s$. To exclude the overlapping of windows, the inequality $s > w$ should be fulfilled. The fragment within one of windows should, first, be mapped onto its complementary counterpart and, then, such complementary fragment can be mapped onto the fragment in the second window by the rules similar to TAMGI. Such complementary mapping retains the mutually complementary nucleotides within windows. Varying $s$-$w$ parameters provides a set of two-parameter complementary mappings.

The pool of $k$-mer motifs reconstructed by TAMGI needs additional analysis. The overlapping shorter motifs can be used for the putative assembly of the longer motifs. Nevertheless, even relatively short motifs with $k \geq 6$ for the step $s = 54$ appeared to be stably reproduced in the genomes of SARS-CoV-2 isolates (Chechetkin and Lobzin, 2020). In particular, the total number of 6-mers at $s = 54$ is 106 for SARS-CoV and 128 for SARS-CoV-2 that is nearly twofold higher than the corresponding number of 6-mers in the randomly reshuffled sequences. This means that the characteristic correlational and quasi-repeating motifs can be used, e.g., for therapeutic targeting, subtyping of viruses or for the assessment of evolutionary divergence between species. The inhomogeneity of motif distribution over the genome and the regions with the highest and the lowest content of motifs may indicate their involvement into molecular mechanisms. The application of the correlational and quasi-periodic motifs to the subtyping of viruses is close to the general discriminant genomic analysis with $k$-mers (Tomović et al., 2006; Ounit et al., 2015).

The short-range correlations of 10–20 nt shown in the inserts to Figs. 3D and 4D facilitate the molecular recognition of long motifs. The near-by positioning of the perfect and mismatch motifs enlarges effectively the recognition region. The proteins bound to such region may consecutively kinetically re-jump from mismatch to perfect motifs.

To sum up, TAMGI method developed in this article is quite general and can be applied to the combined detection/reconstruction of correlational and quasi-periodic motifs in the genomic RNA/DNA sequences. The method provides insight into mechanisms of evolving specific motifs in the genomic sequences. Its scope of applications comprises numerous molecular mechanisms with participation of various functional motifs including quasi-periodic motifs and complete repeats. The method can also be applied to the study of correlations in the genomic sequences.

## Graphical abstract

Genome packaging for SARS-CoV

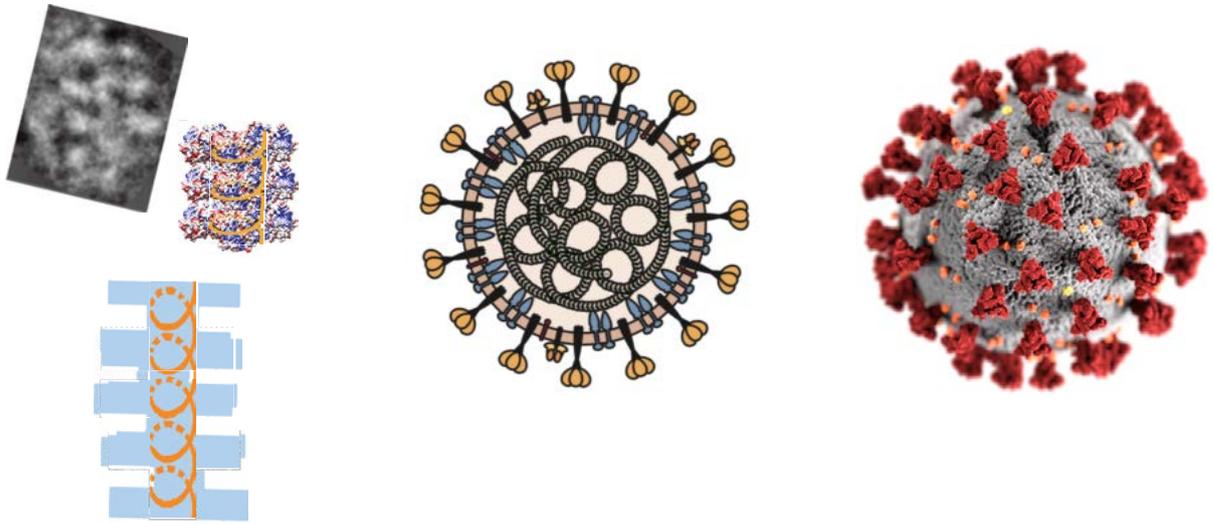

Genome packaging for STMV

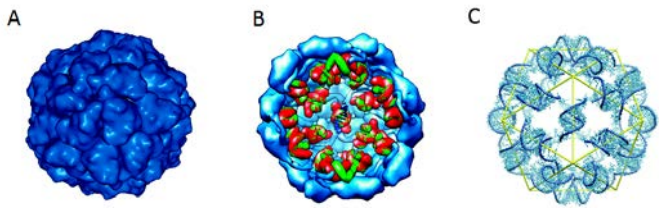

An example of output data for the genome of STMV with the step s=211 and the graphical analysis of TAMGI data on the right

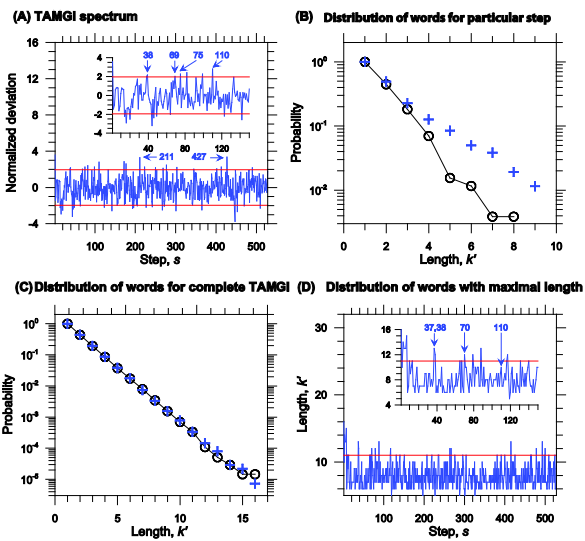

Capsid for PHIX174. The details of genome packaging are unknown

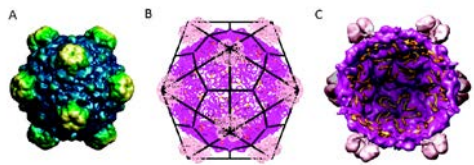